\documentclass[preprint,aps,pra,showpacs]{revtex4}
\usepackage{tabularx}
\usepackage{dcolumn}
\usepackage{graphics}
\usepackage{bm}
\usepackage[dvips]{graphicx}
\usepackage{dcolumn}
\usepackage{tabularx}
\usepackage{amsmath}
\usepackage{amssymb}
\usepackage{color}
\begin{document}
\author{S. M. Sohail Gilani}\affiliation{Centre For High Energy Physics, Punjab
University, Lahore(54590), Pakistan. }
\author{M. Imran Jamil}\affiliation{Department of Physics, School of Science, University of Management and Technology Lahore (54770), Pakistan.}
\author{Bilal Masud}\affiliation{Centre For High Energy Physics, Punjab University,
Lahore(54590), Pakistan.}
\title{$\rho J/\Psi$ Scattering in an Improved Many Body Potential}
\author{Faisal Akram}\affiliation{Centre For High Energy Physics, Punjab University,
Lahore(54590), Pakistan.}
\date{\today}
\begin{abstract}
We calculate the cross-sections for the processes $\rho
J/\psi\rightarrow D^0\bar{D}^0$, $\rho J/\psi\rightarrow
D^0\bar{D}^{0*}$ ($D^{0*}\bar{D}^{0}$) and $\rho J/\psi\rightarrow
D^{0*}\bar{D}^{0*}$ using a QCD-motivated many-body overlap factor to
modify the usual sum of two-body interaction model. The realistic
Cornell potential has been used for pairwise interaction
in the four quark Hamiltonian and noted to give lesser cross-sections as
compared to the quadratic potential. The Resonating group method is employed
along with the Born approximation which decouples its integral equations. It is pointed out that
the additional QCD effect (a gluonic field ovelap factor) result in a
significant suppression in the cross sections as compared to
the more popular sum of two-body interaction.
\end{abstract}
 \maketitle

\section{Introduction}
In the quark potential model the gluonic field energy in quantum
chromodynamics (QCD) is modelled as the potential energy of the
quarks. Its relation with QCD can be seen through the
Born-Oppenheimer approach, used for hadronic physics in
refs.~\cite{morningstar,Nosheen11,Nosheen14,Braaten}. For one pair
of quark and antiquark, Coulombic plus linear or Cornell
form~\cite{Eichten78} for the quark antiquark potential provides a
good fit to the total gluonic field energy for the the lattice
simulations of the gluonic state $\Sigma^+_g$ reported in
refs.~\cite{Morningstar03,Bali,Alexandrou} and many earlier ones.
For a quark antiquark pair, much use has been made of this potential
model to find dynamical implications~\cite{Stephen Godfrey and
Nathan Isgur}. Weinstein and Isgur~\cite{Weinstein 1983, Weinstein
1990} extended this through a sum of such two-body potentials and
argued for the $K\bar{K}$ molecule interpretation of the $f_0(975)$
and $a_0(980)$ mesons.  They variationally optimized a $q^2\bar q^2$
wave function and projected the $q^2\bar q^2$ state onto free $q\bar
q$ wave functions to estimate a relative two meson wave function and
an equivalent meson meson potential. T. Barnes and E. Swanson
further used the sum of pair-wise approach to
calculate~\cite{Barnes1992} $\pi^+\pi^+$, $K^+K^+$ and
$\rho^+\rho^+$ elastic scattering phase shifts and cross sections by
using Born-order quark exchange diagrams (in a nonrelativistic
potential model) which use Gaussian external meson wave functions.
Using the same approach, Barnes et. al. calculated~\cite{Barnes1999}
$BB$ intermeson potentials and scattering amplitudes by taking
Fourier transform of the Born order $T$-matrix element of the
Hamiltonian between two-mesons scattering states. They compare, in
their figures 4 and 5, their nonrelativistic quark model BB
potentials with lattice gauge theory (LGT) results of UKQCD
Collaboration \cite{Michael1999}.

The quark exchange (diagrammatic) approach has been used
~\cite{Martins1995,Barnes2003} for systems with charm and light
quarks, the flavor sectors we address in the present paper. This
model was used by Wong et. al. in refs.~\cite{Wong Swanson Barnes
2000,Wong Swanson Barnes 2001} to calculate the cross sections for
the dissociation of $J/\psi$ and $\psi'$ by $\pi$ and $\rho$ and
later the dissociation cross sections for $J/\psi$, $\psi'$ $\chi$,
$\Upsilon$ and $\Upsilon'$ in collision with $\pi$, $\rho$ and $K$
mesons. It is worthwhile judging if this model agrees to the lattice
simulations of QCD. Such a comparison was made by UKQCD
collaboration and coworkers who noted that the Cornell model
extended in this simple way results in binding energy of two quarks
and two antiquarks square geometry increasing with each quark
separation, whereas the lattice simulations show~\cite{A M Green C
Michael 1993} a decrease. Phenomenologically, the sum of pair wise
interactions results in the long range van der Waal's interaction
between color singlet mesons that contradicts the relevant
experiments. Both of these problems were  remedied~\cite{Masud B
1991} by multiplying the off-diagonal elements in the normalization
and potential energy matrices, in the relevant color basis, through
a space dependent $f$ factor which approaches to 1 in the limit of
all the distances approaching to zero and to zero in the limit of
large separations.

The actual form of this gluonic field overlap factor $f$ and fit of
its parameter(s) to the relevant lattice simulations was improved
through refs.~\cite{A M Green C Michael 1992,B Masud 1995,Petrus
Pennanen 1997,A M Green J Koponen 1999}. Meanwhile, ref.~\cite{W R
Thomas1990} calculated terms in the effective Hamiltonian for quark
(and antiquark) kinetic energy operators. This allowed incorporating
the quark motion, with the $f$ factor appearing in the off-diagonal
elements of the kinetic energy matrices as well. The $f$ factor
affects the linear independence of the color configurations. But
still a two-states basis was found~\cite{Petrus Pennanen 1997,A M
Green J Koponen 1999} to be sufficient to well model the relevant
computer simulations; we use $2\times 2$ matrices.  We have used the
simplest (Gaussian) form of $f$ that is used in  ref.~\cite{A M
Green J Koponen 1999} along with the best fitted value of $k_f$
chosen in it. But, as in refs.~\cite{Masud B 1991,Imran Jameel
2011}, we multiply it by the matrix elements of the actual SU(3)
color matrices. In these and related calculations~\cite{Masud B
1994, Imran Jameel 2017} for meson-meson cross-sections and
bindings, resonating group method (mentioned below in section II)
has been employed to use the wave function of a single cluster of a
quark antiquark pair. These wave functions are taken to be those of
the quadratic confinement; each potential energy is also quadratic
in the four papers.  In the present paper we fit the parameters of a
Gaussian form to the numerically calculated eigenfunctions of the
realistic Cornell potential for each cluster, and in the multiquark
Hamiltonian we have Cornell potential for each pair-wise
interaction. Here we incorporate the spin and flavor dependence and
use Born approximation to report the cross sections for the
processes
 $\rho J/\psi\rightarrow D^0\bar{D}^0$,
$\rho J/\psi\rightarrow D^0\bar{D}^{0*}$ or $D^{0*}\bar{D}^{0}$,
$\rho J/\psi\rightarrow D^{0*}\bar{D}^{0*}$,
$D^0\bar{D}^0\rightarrow \rho J/\psi$, $D^{0*}\bar{D}^{0}$ or
$D^0\bar{D}^{0*}\rightarrow \rho J/\psi$ and
$D^{0*}\bar{D}^{0*}\rightarrow \rho J/\psi$.
The values of T-Matrix elements (phase shifts)\cite{Imran Jameel
2011} are much less than 1 radian which indicates the validity of
the Born approximation.
It is expected since ref.~\cite{Barnes1992} has in it that ``~There
is much circumstantial evidence that high-order diagrams are
relatively unimportant in hadron spectroscopy and in low energy
scattering and decays (excluding the $u\bar{u} \leftrightarrow
d\bar{d}$ mixing)" and Born approximation means neglecting these
high-order diagrams.

We report a comparison of incorporating the QCD-motivated improved
overlap factor $f$ with the more popular sum of two-body approach,
along with those of almost a full use of the realistic Cornell
potential and of replacing this by a computationally convenient
quadratic potential. A quadratic potential is frequently
used~\cite{Suzuki1983,Burger 1989,Schroder1986, 0112034v2} in the
hadronic physics including the nucleon-nucleon interaction. In the
old Isgur-Karl model, still used in ref.~\cite{0906.0699v1}, the
leading order quark antiquark interaction is quadartic. Ding Xiaonan
in ref.~\cite{DingXiaonan1988} used the quadratic and linear
potentials for baryon baryon interaction and got similar phase
shifts. Qualitative properties of QCD Benzene found by assuming
quadratic confinement in ref.~\cite{0610390v3} then found the
uncertainty in the masses of benzene like structure with three
diquark and colour octet states when compared with the linear
confinement. Zahra Ghalenovi et. al. in ref.~\cite{1507.03345v1}
used both linear and quadratic confinement along with Coulomb like
term in the potential to calculate mass spectra of heavy and light
scalar tetraquarks modelled as bound states of point-like diquark
and diantiquark. They also introduced spin spin, spin isospin and
isospin isospin interactions during their calculations and found the
compatibility with mild influence of the confining potential.

The paper is organized as follows: In Sec. \ref{sec2}, we write the total Hamiltonian of the four quark
system in the sum of two body approach. And then we write the total
state vector of diquark diantiquark system by using the adiabatic
approximation. By using the Resonating Group Method we write the
resultant integral equations. In Sec. \ref{coupled equations}, we
decouple the two coupled equations by using the Born approximation
then we solve them. At the end of this section we describe the
corresponding changes in the two equations for different studied
processes. In Sec. \ref{crosssections} we describe the formalism to
obtain the cross sections. In Sec. \ref{parameters} we describe the
phenomenological fitting of the parameters. In Sec. \ref{conclusion}
we give the obtained cross sections for different processes. Further
more we compare the cross sections for two different potentials. We
also compare the cross sections for the Gaussian form of $f$ with
the simple sum of two body approach as well. A comparison with other
works is also given there.

\section{The Hamiltonian and Gluonic Basis}\label{sec2}
In the sum of two body potential model, the total Hamiltonian of the four quarks system
is defined as
\begin{equation}
H=\sum_{i=1}^{\bar{4}}\bigg[m_i+\frac{P_i^2}{2m_i}\bigg]+\sum_{i<j}(v_{ij}+H_{hyp}^{ij})\textbf{F}_i.\textbf{F}_j,
\end{equation}
where
\begin{equation}
H_{hyp}^{ij}=-\frac{8\pi\alpha_s}{3m_im_j}\textbf{S}_i.\textbf{S}_j\delta(r_{ij}).
\end{equation}
For the \textit{i}th particle, $m_i$ is constituent quark mass, $\textbf{S}_i$ is the set of spin matrices,
and $\mathbf{F}_{i}$ is the set of $SU(3)_c$ matrices. The components of $\textbf{F}$ are
$\frac{\lambda_{a}}{2}$ for a quark and $-\frac{\lambda_{a}^*}{2}$ for an anti quark, with $a=1,2,3,...,8$. The $v_{ij}$ is inter-quark Cornell potential, which is
\begin{equation}v(\textbf{r}_{ij})=
v_{ij}=\frac{\alpha_s}{r_{ij}}-\frac{3}{4}b~r_{ij}+c~~~~~~~~ \text{
with } i,j=1,2,\bar{3},\bar{4}.\label{vij}
\end{equation}
Here $b$ is string tension, $\alpha_{s}$ is strong coupling
constant and c is the self energy constant. We incorporate the energy and hence
flavor dependence of the strong coupling and self energy constant by taking their values to be
 $\alpha_{s1}$ and $c_1$ for the mesons having quarks content $u\bar{c}$, and $\alpha_{s2}$, $c_2$ and
$\alpha_{s3}$, $c_3$ for the $u\bar{u}$ and $c\bar{c}$ clusters respectively.
The phenomenological fitting of these parameters is given in Sec. \ref{parameters}. In the same way, where we use for a
comparison the quadratic potential
\begin{equation}v(\textbf{r}_{ij})=
v_{ij}=Cr_{ij}^2+\bar{C}\label{vqij}
\end{equation}
the three $C's$ along with $\bar{C}$ are phenomenologically fitted as in Sec. \ref{parameters}.

\begin{figure}
\includegraphics[scale=.60,angle=-0]{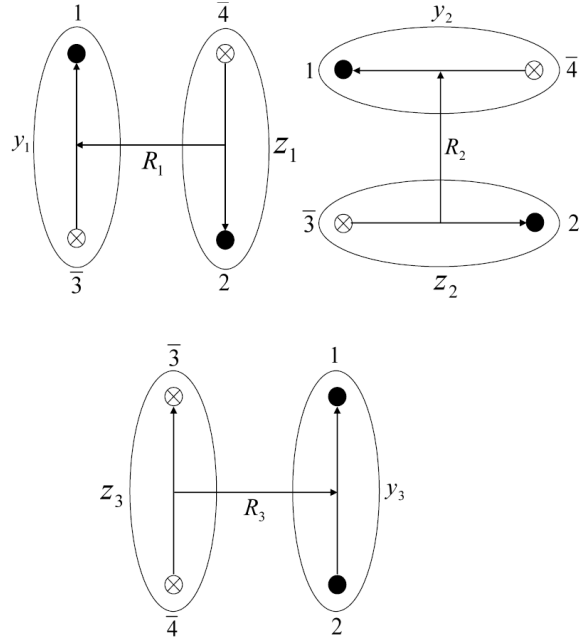}
\caption{Different topologies of diquark and diantiquark system. }
\label{fig1}
\end{figure}

The total state vector of our diquark diantiquark system can be
written as
\begin{equation}
|\psi(\textbf{r}_1,\textbf{r}_2,\textbf{r}_{\bar3},\textbf{r}_{\bar4};g)\rangle
=\sum_{k=1}^2|k\rangle_g|k\rangle_s|k\rangle_f|
\psi(\textbf{r}_1,\textbf{r}_2,\textbf{r}_{\bar3},\textbf{r}_{\bar4})\rangle.
\end{equation}
We have neglected the third topology of Fig. \ref{fig1} because the lattice
energies in Table IV of Ref. \cite{Green
Lukkarinen Pennanen 1996} are essentially unaffected by this truncation.
\noindent $\{|k\rangle_g\}$ is the gluonic basis which reduces to color
basis $\{|1_{1\bar 3}1_{2\bar 4}\rangle,|1_{1\bar 4}1_{2\bar 3}\rangle \}$ in the weak coupling limit. The spin basis is composed of
\begin{eqnarray}
\mathrm|1\rangle_s=\left\lbrace \begin{array}{lll}
|P_{1\bar{3}}P_{2\bar{4}}\rangle_0~~~~~~~~~~ \textrm{for}~ D^0\bar{D}^0 \\
|V_{1\bar{3}}V_{2\bar{4}}\rangle_{0,1\hspace{.03in}\text{or}\hspace{.03in}2}~~~~~~~~~~ \textrm{for}~
D^{0*}\bar{D}^{0*}~
 \\
|P_{1\bar{3}}V_{2\bar{4}}\rangle_1~~~~~~~~~~\mathrm{for}~D^{0}\bar{D}^{0*}
\end{array} \right. \label{1s},
\end{eqnarray}
\begin{eqnarray}
\mathrm|2\rangle_s= |V_{1\bar{4}}V_{2\bar{3}}\rangle_{0,1\hspace{.03in}\text{or}\hspace{.03in}2}~~~~~~~~~~~~
\textrm{for}~\rho J\psi\label{2s}.
\end{eqnarray}
These spin states are explicitly defined
with their overlaps and $\textbf{S.S}$ matrix elements in the Appendix A. The flavor contents of first and 2nd channel are
\begin{equation}
|1\rangle_f=|u\bar{c}\rangle_{1\bar 3}|c\bar{u}\rangle_{2\bar
4}\label{1f},
\end{equation}
\begin{equation}
|2\rangle_f=\frac{1}{\sqrt{2}}|u\bar{u}-d\bar{d}\rangle_{1\bar
4}|c\bar{c}\rangle_{2\bar 3}\label{2f},
\end{equation}
which gives
\begin{equation}
{}_f\langle1|2\rangle_f={}_f\langle2|1\rangle_f=\frac{1}{\sqrt{2}}\label{flavor
overlap}.
\end{equation}
The four 3-vectors $\textbf{r}_1,\textbf{r}_2,\textbf{r}_{\bar3}$
and $\textbf{r}_{\bar4}$ can be replaced by their linear combinations as
 $\textbf{R}_c$ (center of mass) along with $\textbf{R}_k$, $\textbf{y}_k$ and $\textbf{z}_k$ shown in Fig.~\ref{fig1}. This means
 \begin{equation}
\textbf{R}_1=\frac{(\textbf{r}_1-\textbf{r}_{\bar4})+r_c(\textbf{r}_{\bar3}-\textbf{r}_2)}{1+r_c}\label{R1},
\end{equation}
where $r_c=\frac{m_c}{m_u}$. $\textbf{R}_2$ and $\textbf{R}_3$ are
similarly defined. As in the resonating group method, we factorize
the dependence on the three relative vectors as a product of the
inter-cluster ($\textbf{R}_k$) dependence and the intra-clusters (
$\textbf{y}_k$ and $\textbf{z}_k$) dependence. This makes the total
state vector as
\begin{equation}
|\psi(\textbf{r}_1,\textbf{r}_2,\textbf{r}_{\bar3},\textbf{r}_{\bar4};g)\rangle
=\sum_{k=1}^2|k\rangle_g|k\rangle_s|k\rangle_f
\psi_c(\textbf{R}_c)\chi_k(\textbf{R}_k)\xi_k(\textbf{y}_k)\zeta_k(\textbf{z}_k)\label{state
vector}.
\end{equation}
\noindent $\xi_k(\textbf{y}_k)$ and $\zeta_k(\textbf{z}_k)$ are intra-cluster
wave functions which we take to be in Gaussian form as following
\begin{equation}
\xi_k(\textbf{y}_k)=\frac{1}{(2\pi
d_{k1}^2)^{3/4}}exp\bigg(\frac{-\textbf{y}_k^2}{4d_{k1}^2}\bigg),
\end{equation}
\begin{equation}
\zeta_k(\textbf{z}_k)=\frac{1}{(2\pi
d_{k2}^2)^{3/4}}exp\bigg(\frac{-\textbf{z}_k^2}{4d_{k2}^2}\bigg),
\end{equation}
where $d_{k1}$ and $d_{k2}$ are corresponding mesonic sizes.

Substituting Eq. (\ref{state
vector}) in \begin{equation}
\langle\delta\Psi|H-E|\Psi\rangle=0
\end{equation}
\noindent and taking linearly independent variations only in
$\chi_k$ factor, along with performing the trivial $\textbf{R}_c$ integration using, say, a box normalization, we get
\begin{eqnarray}
\sum_{l=1}^2\int d^3\textbf{y}_k
d^3\textbf{z}_k\xi_k(\textbf{y}_k)\zeta_k(\textbf{z}_k){\ _f\langle
k|}{\ _s\langle k|}{\ _g\langle
k|}H-E|l\rangle_g|l\rangle_s|l\rangle_f
\chi_l(\textbf{R}_l)\xi_l(\textbf{y}_l)\zeta_l(\textbf{z}_l)=0\label{equation},
\end{eqnarray}
which defines two integral equations for $k$=1 and 2. The matrix elements of potential energy part of the
Hamiltonian in our spin basis are given by
\begin{equation}
_s\langle k|
V(f)|l\rangle_s=\sum_{i<j}\textbf{F}_i.\textbf{F}_j(f)(V_{ij})_{k,l},\label{eq27}
\end{equation}
with
\begin{equation}
(V_{ij})_{k,l}=v_{ij}\ _s\langle
k|l\rangle_s-\frac{8\pi\alpha_s}{3m_im_j}\delta(r_{ij})\ _s\langle
k|\textbf{S}_i.\textbf{S}_j|l\rangle_s.
\end{equation}
Now using the matrix elements of $\textbf{F}_i.\textbf{F}_j$
 operator in the colour basis \cite{Weinstein 1990}
in the potential energy matrix, defined in Eq. (\ref{eq27}) above, in gluonic basis modified by $f$ model is given by
 \begin{equation}
V(f)=\left(
    \begin{array}{cc}
      -\frac{4}{3}(V_{1\bar3}+V_{2\bar4})_{1,1} & \frac{4f}{9}(V_{12}+V_{\bar3\bar4}-V_{1\bar3}-V_{2\bar4}-V_{1\bar4}-V_{2\bar3})_{1,2} \\
      \frac{4f}{9}(V_{12}+V_{\bar3\bar4}-V_{1\bar3}-V_{2\bar4}-V_{1\bar4}-V_{2\bar3})_{2,1} & -\frac{4}{3}(V_{1\bar3}+V_{2\bar4})_{2,2} \\
    \end{array}
  \right)\label{V}.
\end{equation}
Here, $f$, is defined as \cite{A M Green J Koponen 1999}
\begin{equation}
f=\text{exp}(-b k_f\sum_{i<j}r_{ij}^2)\label{f},
\end{equation}
where $k_f$ is a parameter fitted to minimize the model and
simulation binding energies. The color basis is non-orthogonal. Thus
there is a non-trivial overlap matrix which is modified by the $f$-model to become
 \begin{equation}
N(f)\equiv\{_g\langle k|l\rangle_g\}=\left(
    \begin{array}{cc}
      1 & \frac{f}{3} \\
      \frac{f}{3} & 1 \\
    \end{array}
  \right).\label{Nf}
\end{equation}
The $k,l$ matrix element of the kinetic energy matrix (in a Hermitian form) in the gluonic basis  is
\begin{equation}
_g\langle
k|K|l\rangle_g=N(f)^{1/2}_{k,l}\bigg(\sum_i-\frac{\nabla_i^2}{2m_i}\bigg)N(f)^{1/2}_{k,l}\label{Kf}.
\end{equation}
\section{Writing the Coupled Equations}\label{coupled equations}
The total spin for each of $\rho J/\Psi\rightarrow D^0\bar{D^0}$ and
$D^0\bar{D^0}\rightarrow \rho J/\Psi$ is zero. For the first channel we take $k=1$ in Eq. (\ref{equation}).
Our Hamiltonian is identity in the flavor space, so we use flavor overlaps from Eq.
(\ref{flavor overlap}). The spin overlaps and \textbf{S.S} matrix
elements in spin basis are given in
Appendix \ref{spin}. We insert the elements of P.E., Normalization and K.E  Matrices from Eqs. (\ref{V}, \ref{Nf} and \ref{Kf}) in Eq. (\ref{equation}) and use Born approximation i.e. non-interacting
$\chi_1(\textbf{R}_1)=\sqrt{2/\pi}e^{i\textbf{P}_1.\textbf{R}_1}$ to
decouple the two coupled equations (see Eq. (\ref{equation})). Performing the Fourier transform with respect to $\textbf{R}_1$ we get
\begin{eqnarray}
(k_1+k_2{\textbf{P}_1}^2-E+2m_u(1+r_c))\chi_1(\textbf{P}_1)=h_{12}\label{CE1},\hspace{1in} \text{where}
\end{eqnarray}
\begin{eqnarray}
h_{12}=\frac{1}{(2\pi)^{\frac{3}{2}}}\frac{1}{\sqrt{2}}\int
d^3\textbf{y}_1d^3\textbf{z}_1d^3\textbf{R}_1\xi_1(\textbf{y}_1)\zeta_1(\textbf{z}_1)
\sqrt{f}\bigg\{-\frac{4}{9}(V_c-V_{hyp})\sqrt{f}e^{i
\textbf{P}_2.\textbf{R}_2}\xi_2(\textbf{y}_2)\nonumber\\
\zeta_2(\textbf{z}_2)+\mathcal{K}_{12}-\mathcal{N}_{12}\bigg\}\sqrt{\frac{2}{\pi}}e^{i
\textbf{P}_1.\textbf{R}_1}.\label{h12}~~~~~~~~~~~~~~~~~~~~~~~~~~~~~~~~~~~~~~~~
\end{eqnarray}
 In Eq. (\ref{h12}) $V_c$, $V_{hyp}$, $\mathcal{K}_{12}$ and $\mathcal{N}_{12}$ are defined as
\begin{eqnarray}
V_c=\frac{-\sqrt{3}}{2} \bigg\{\alpha_{s1}\bigg(\frac{1}{y_3}
+\frac{1}{z_3}-\frac{1}{y_1}
-\frac{1}{z_1}\bigg)-\frac{\alpha_{s2}}{y_2}-\frac{\alpha_{s3}}{z_2}
\nonumber\\
-\frac{3}{4}b_s(y_3+z_3-y_1-z_1-y_2-z_2)-c_2-c_3\bigg\}\label{Vc}.
\end{eqnarray}
\begin{eqnarray}
V_{hyp}=\frac{\sigma^3}{\sqrt{3\pi}}
\bigg\{\alpha_{s1}\bigg(\frac{e^{-\sigma^2\textbf{y}_3^2}}{m_u
m_c}+\frac{e^{-\sigma^2\textbf{z}_3^2}}{m_c m_u}
-\frac{3e^{-\sigma^2\textbf{y}_1^2}}{m_u m_c}
-\frac{3e^{-\sigma^2\textbf{z}_1^2}}{m_c m_u}\bigg)
+\frac{\alpha_{s2}e^{-\sigma^2\textbf{y}_2^2}}{m_u m_u}
+\frac{\alpha_{s3}e^{-\sigma^2\textbf{z}_2^2}}{m_c
m_c}\bigg\}\label{Vhyp}.
\end{eqnarray}
\begin{eqnarray}
\mathcal{K}_{12}=\frac{1}{2\sqrt{3}} \big(-\frac{1}{2m_u}
\big)\big[\xi_2(\textbf{y}_2)\zeta_2(\textbf{z}_2)
f_2\nabla_{\textbf{R}_2}^2(\sqrt{f}e^{i
\textbf{P}_2.\textbf{R}_2})+e^{i
\textbf{P}_2.\textbf{R}_2}\zeta_2(\textbf{z}_2)g_2\nabla_{\textbf{y}_2}^2(\sqrt{f}\xi_2(\textbf{y}_2))
\nonumber\\
+e^{i \textbf{P}_2.\textbf{R}_2}
\xi_2(\textbf{y}_2)h_2\nabla_{\textbf{z}_2}^2(\sqrt{f}\zeta_2(\textbf{z}_2))\big],~~~~~~~~~~~~~~~~~~~~~~~~~~~~~~~~~~~~~~~~
\end{eqnarray}
where $f_2=\frac{1}{2}(\frac{1+r_c}{r_c})$, $g_2=2$, and
$h_2=\frac{2}{r_c}$.
\begin{eqnarray}
\mathcal{N}_{12}=(E-2m_u(1+r_c))\big(\frac{1}{2\sqrt{3}}\big)
\sqrt{f}e^{i \textbf{P}_2.\textbf{R}_2}
\xi_2(\textbf{y}_2)\zeta_2(\textbf{z}_2).
\end{eqnarray}
We chose $z$-axis along $\textbf{P}_1$ and $x$-axis in
the plane containing $\textbf{P}_1$ and $\textbf{P}_2$. Naming the angle between $\textbf{P}_1$ and $\textbf{P}_2$ as $\theta$, we get $P_{2x}=P_2\text{Sin}\theta$, $P_{2z}=P_2\text{Cos}\theta$,
$P_{2y}=0$, along with $P_{1x}=P_{1y}=0$, $P_{1z}=P_1$.

Similarly, for the second channel, we take $k=2$ in Eq.
(\ref{equation}) and apply the same procedure as for Eq.(\ref{CE1}) to get
 \begin{eqnarray}
(k_3+k_4{\textbf{P}_2}^2-E+2m_u(1+r_c))\chi_2(\textbf{P}_2)=h_{21}\label{CE2},\hspace{1in}\text{where}
\end{eqnarray}
\begin{eqnarray}
h_{21}=\frac{1}{(2\pi)^{\frac{3}{2}}} \frac{1}{\sqrt{2}} \int
d^3\textbf{y}_2d^3\textbf{z}_2d^3\textbf{R}_2\xi_2(\textbf{y}_2)\zeta_2(\textbf{z}_2)
\sqrt{f}\bigg\{-\frac{4}{9}(V_c-V_{hyp})\sqrt{f}e^{i
\textbf{P}_1.\textbf{R}_1}\nonumber\\
\xi_1(\textbf{y}_1)\zeta_1(\textbf{z}_1)+\mathcal{K}_{21}-\mathcal{N}_{21}\bigg\}\sqrt{\frac{2}{\pi}}e^{i
\textbf{P}_2.\textbf{R}_2}\label{h21}.~~~~~~~~~~~~~~~~~~~~~~~~
\end{eqnarray}
$V_c$ and $V_{hyp}$ are same as in Eq. (\ref{CE1}). $\mathcal{K}_{21}$ and $\mathcal{N}_{21}$ are defined as
\begin{eqnarray}
\mathcal{K}_{21}=\frac{1}{2\sqrt{3}}(-\frac{1}{2m_u}
\big)\sqrt{f}[\xi_1(\textbf{y}_1)\zeta_1(\textbf{z}_1)
f_1\nabla_{\textbf{R}_1}^2(\sqrt{f}e^{i\textbf{P}_1.\textbf{R}_1})+e^{i\textbf{P}_1.\textbf{R}_1}
\zeta_1(\textbf{z}_1)g_1\nonumber\\
\nabla_{\textbf{y}_1}^2(\sqrt{f}\xi_1(\textbf{y}_1))
+e^{i\textbf{P}_1.\textbf{R}_1}\xi_1(\textbf{y}_1)h_1\nabla{\textbf{z}_1}^2
(\sqrt{f}\zeta_1(\textbf{z}_1))],
\end{eqnarray}
where $f_1=\frac{2}{1+r_c}$, $g_1=\frac{1+r_c}{r_c}$ and
$h_1=\frac{1+r_c}{r_c}$.
\begin{eqnarray}
\mathcal{N}_{21}=(E-2m_u(1+r_c))\frac{1}{2\sqrt{3}}\sqrt{f}e^{i\textbf{P}_1.\textbf{R}_1}
\xi_1(\textbf{y}_1)\zeta_1(\textbf{z}_1).
\end{eqnarray}
Above is for the processes $\rho J/\Psi\rightarrow D^0\bar{D^0}$ and
$D^0\bar{D^0}\rightarrow \rho J/\Psi$. Due to different spin states the hyperfine term for the
process $\rho J/\psi\rightarrow D^{0}\bar D^{0*}$ or $D^{0}\bar
D^{0*}\rightarrow \rho J/\psi$ is given instead by
\begin{equation}
V_{hyp}=-\sqrt{\frac{2}{3}}\frac{\sigma^3}{\sqrt{3\pi}}
\bigg\{\alpha_{s1}\bigg(\frac{e^{-\sigma^2\textbf{y}_3^2}}{m_um_c}
+\frac{e^{-\sigma^2\textbf{z}_3^2}}{m_cm_u}-\frac{3e^{-\sigma^2\textbf{y}_1^2}}{m_um_c}
+\frac{e^{-\sigma^2\textbf{z}_1^2}}{m_cm_u}\bigg)
+\frac{\alpha_{s2}e^{-\sigma^2\textbf{y}_2^2}}{m_u
m_u}+\frac{\alpha_{s3}e^{-\sigma^2\textbf{z}_2^2}}{m_c m_c}\bigg\},
\end{equation}
and all the remaining terms of Eqs. (\ref{h12} and \ref{h21})
are multiplied by a factor $-\sqrt{2/3}$, because the corresponding spin overlap
of the above mentioned processes is $1/\sqrt{2}$.

For the process $\rho J/\psi\rightarrow D^{0*}\bar D^{0*}$ or
$D^{0*}\bar D^{0*}\rightarrow \rho J/\psi$ the hyperfine term differs with the total spin.
For total spin zero this becomes
\begin{equation}
V_{hyp}=\frac{1}{\sqrt{3}}\frac{\sigma^3}{\sqrt{3\pi}}
\bigg\{\alpha_{s1}\bigg(\frac{5e^{-\sigma^2\textbf{y}_3^2}}{m_um_c}
+\frac{5e^{-\sigma^2\textbf{z}_3^2}}{m_cm_u}+\frac{e^{-\sigma^2\textbf{y}_1^2}}{m_um_c}
+\frac{e^{-\sigma^2\textbf{z}_1^2}}{m_cm_u}\bigg)
+\frac{\alpha_{s2}e^{-\sigma^2\textbf{y}_2^2}}{m_u
m_u}+\frac{\alpha_{s3}e^{-\sigma^2\textbf{z}_2^2}}{m_c m_c}\bigg\},
\end{equation}
and all the remaining terms of Eqs. (\ref{h12} and
\ref{h21}) are multiplied by a factor $1/\sqrt{3}$, because the corresponding spin
overlap of the above mentioned processes is $-1/2$. For total spin 2
the hyperfine term becomes
\begin{equation}
V_{hyp}=\frac{2}{\sqrt{3}}\frac{\sigma^3}{\sqrt{3\pi}}
\bigg\{\alpha_{s1}\bigg(\frac{e^{-\sigma^2\textbf{y}_3^2}}{m_um_c}
+\frac{e^{-\sigma^2\textbf{z}_3^2}}{m_cm_u}-\frac{e^{-\sigma^2\textbf{y}_1^2}}{m_um_c}
-\frac{e^{-\sigma^2\textbf{z}_1^2}}{m_cm_u}\bigg)
-\frac{\alpha_{s2}e^{-\sigma^2\textbf{y}_2^2}}{m_u
m_u}-\frac{\alpha_{s3}e^{-\sigma^2\textbf{z}_2^2}}{m_c m_c}\bigg\},
\end{equation}
and all the remaining terms of Eqs. (\ref{h12} and \ref{h21}) are multiplied by a factor $-2/\sqrt{3}$, because the corresponding spin overlap of the above mentioned processes is $1$.

When we use quadratic potential then only potential part of the
integral Eqs. (\ref{CE1} and \ref{CE2}) is changed. That is $V_c$
is replaced by $V_q$, where
\begin{equation}
V_q=\frac{-\sqrt{3}}{2}\bigg\{C_2\bigg(\textbf{y}_3^2+\textbf{z}_3^2-\textbf{y}_1^2-\textbf{z}_1^2\bigg)
-C_3\textbf{y}_2^2-C_1\textbf{z}_2^2-2\bar{C}\bigg\}.
\end{equation}
Here, $C_1$ is a constant of inter-quark potential (see Eq. \ref{vqij})  of $c\bar c$ mesons, $C_2$ is constant for
$c\bar u $ or $u\bar c$ mesons and $C_3$ is constant for $u\bar u$
mesons. When we use quadratic potential in the coupled equations
(\ref{equation}), the spin averaging is used to fit the parameters
given in Sec. \ref{parameters} and hence the hyperfine term is neglected.
\section{Finding the Cross-Sections}\label{crosssections}
The $T$ matrix elements can be read off from Eqs. (\ref{CE1}) and
(\ref{CE2}). These would be proportional to the coefficients
($h_{12}$ and $h_{21}$) of the non relativistic Green operators
$-1/\vartriangle_1(P_1)$ and $-1/\vartriangle_2(P_2)$, where
\begin{equation}
\vartriangle_1(P_1)=(k_1+k_2{\textbf{P}_1}^2-E+2m_u(1+r_c))\hspace{1in}\text{and}
\end{equation}
\begin{equation}
\hspace{-1.2in}\vartriangle_2(P_2)=(k_3+k_4{\textbf{P}_2}^2-E+2m_u(1+r_c)).
\end{equation}
 That is, \cite{Masud B 1994}
\begin{equation}
T_{12}=2\mu_{12}\frac{\pi}{2}P_2\sqrt{\frac{v_1}{v_2}}h_{12}\label{Tij}
\end{equation}
\begin{equation}
T_{21}=2\mu_{34}\frac{\pi}{2}P_1\sqrt{\frac{v_2}{v_1}}h_{21}\label{Tji},
\end{equation}
where $i,j=1,2$ and $P_j$ are relative momenta of two clusters for
channel 1 and 2 defined as
\begin{equation}
P_1=\sqrt{2\mu_{12}(E-M_1-M_2)}\hspace{1in}\text{and}\label{p1}
\end{equation}
\begin{equation}
\hspace{-1.3in}P_2=\sqrt{2\mu_{34}(E-M_3-M_4)}.\label{p2}
\end{equation}
Here $M_i,i=1,2,3,4$ are corresponding meson masses and $\mu_{ij}$
are reduce masses of corresponding mesons, $v_1=P_1/\mu_{12}$,
$v_2=P_2/\mu_{34}$.
Finally we get the spin averaged cross-sections by using the following equation
\cite{weinberg}
\begin{equation}
\sigma_{ij}=\frac{4\pi}{P_j^2}\sum_J\frac{2J+1}{(2s_1+1)(2s_2+1)}|T_{ij}|^2,\label{sigmaij}
\end{equation}
where $J$ is the total spin of the two outgoing mesons and $s_1$ and
$s_2$ are the spins of the two incoming mesons.

\section{Fitting the Parameters}\label{parameters}
The parameters for Cornell Potential i.e $\alpha_s$ and $c$ are
fitted by finding minimum Chi square between the masses taken from
PDG and the mass spectrum generated using Cornell model in the quark
potential model for the mesons $\rho_0, b_1, a, D^0, D^{0*}, D_1,
D_2^*, \eta_c, J/\psi, h_c$ and $\chi_c$ . The fitted parameters are
$\alpha_{s1}=0.38$, $\alpha_{s2}=\alpha_{s3}=0.5$, $c_1=0.732$,
$c_2=0.692$, and $c_3=0.612$. The values $m_u=0.345$, $m_c=1.931$,
$k_f=0.075$, $b_s=0.18$, and $\sigma=0.897$ are taken from Refs.
\cite{Wong Swanson Barnes 2000,Wong Swanson Barnes 2001,A M Green J
Koponen 1999}. In quadratic potential $\omega$, $m_u$ and $m_c$ are
fitted by minimizing the chi-square
$\chi^2=\sum_i((E_{i}-\tilde{E_{i}})/\tilde{E_{i}})^2$, where
$\tilde{E_{i}}$ are the spin averaged experimental masses \cite{PDG}
 and $E_{i}$ are the masses obtained by
$E_{i}=(\omega/2)(4n+2l+3)+m_0$\cite{Zettili}, that is 3-d S.H.O.
energy and the rest mass energy. Here $i$ labels the different
mesons. We have used the following spectrum to fit the parameters
$\eta_c, J/\psi, \chi_{c0}, \chi_{c1}, \chi_{c2}, D^0, D^{0*}, D_1,
D_2, \rho, \omega, a_1, a_2, b_1, \pi(1300), \rho_3$ and $\pi_2$.
After getting $\omega$ the constants and mesons sizes are obtained
by using $C=-(3/16)(2\mu \omega^2$) and
$d=\sqrt{1/2\mu\omega}$\cite{Masud B 1991,Masud B 1994}. Hence the
fitted parameters are $m_u=0.1065$, $m_c=1.2877$, $C_1=-0.0273$,
$C_2=-0.00615$, $C_3=-0.002995$ and $\bar C=0$. The resulting meson
sizes for both quadratic and Cornell potential are given in Table
\ref{table1}, along with the meson masses used in the fitting taken
from PDG.
\begin{table}[!htb]
\begin{center}
\begin{tabular}{|c|c|c|c|}
  \hline
  &  &  Cornell potential &  Quadratic Potential \\
  \hline
  Meson & Mass(GeV) & d$(GeV^{-1})$ & d$(GeV^{-1})$  \\
  \hline
  $\rho$ & 0.77549 & 2.3827 & 4.9241 \\ \hline
  $D^0/\overline{D}^0$ & 1.86483 & 1.6543 & 3.5288 \\ \hline
  $D^{0*}/\overline{D}^{0*}$ & 2.00698 & 1.9238 & 3.5288 \\ \hline
  $J/\psi$ & 3.096916 & 0.9947 & 1.5199 \\
  \hline
\end{tabular}
\caption{Mesons Masses and their corresponding sizes for Cornell and
quadratic Potentials.}\label{table1}
\end{center}
\end{table}
\section{Results and Conclusions}\label{conclusion}
\begin{figure}[!h]
\includegraphics[scale=.90,angle=-0]{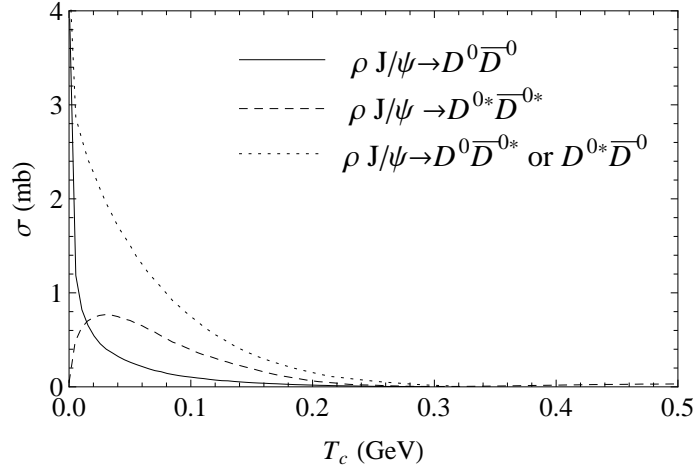}
\caption{Graph between Cross sections and $T_c$ for Cornell
potential with $f=1$.} \label{cornell}
\end{figure}

\begin{figure}[!h]
\includegraphics[scale=.90,angle=-0]{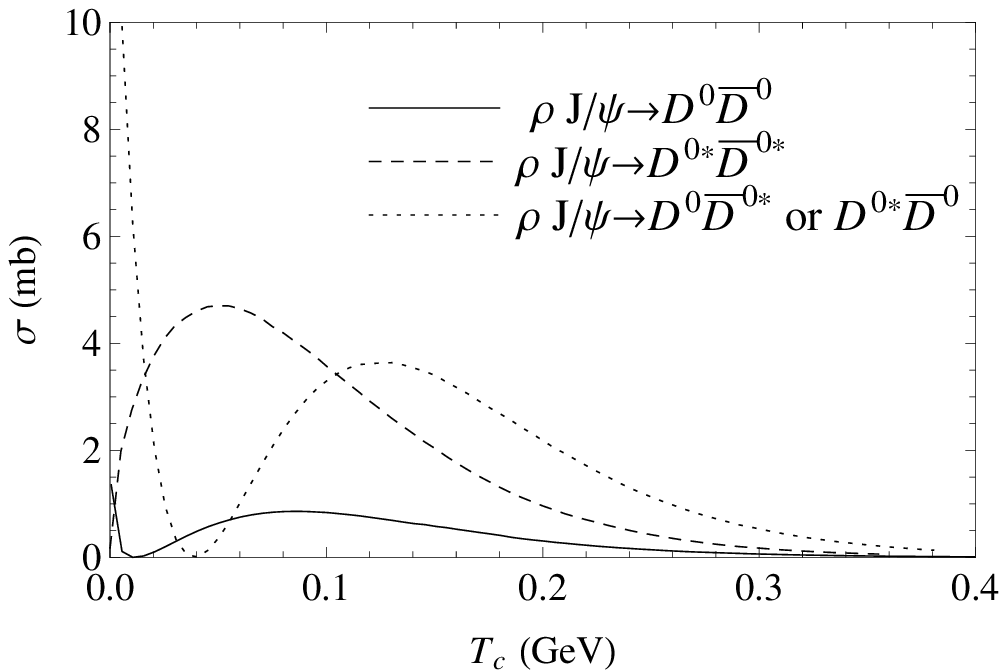}
\caption{Graph between Cross sections and $T_c$ for Quadratic
potential with $f=1$.} \label{quadratic}
\end{figure}

\begin{figure}[!h]
\includegraphics[scale=.90,angle=-0]{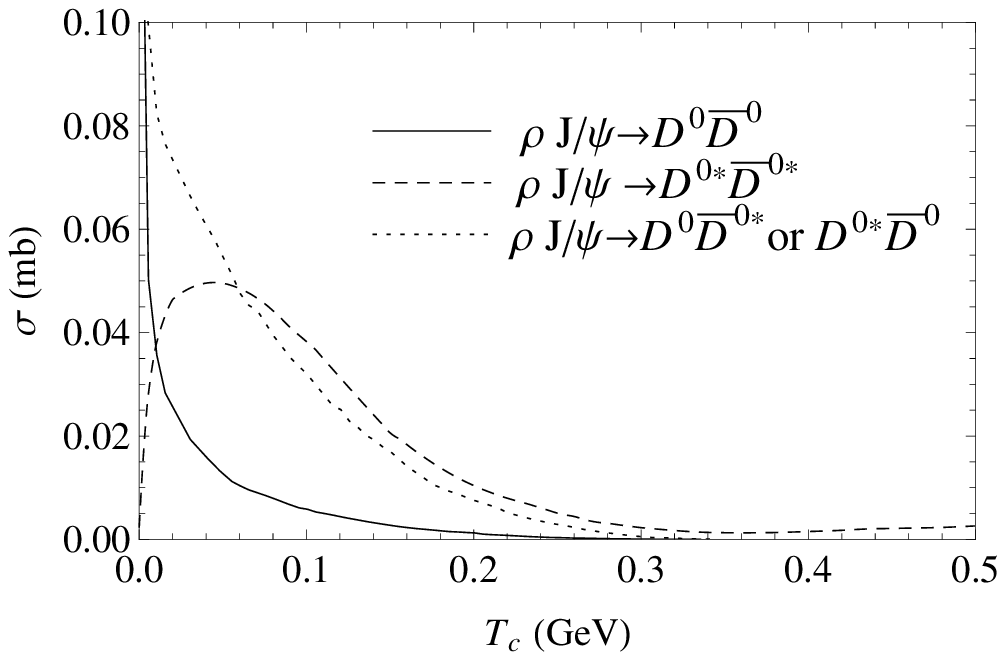}
\caption{Graph between Cross sections and $T_c$ for Cornell
potential with Gaussian form of $f$.} \label{cornellf}
\end{figure}

\begin{figure}[!h]
\includegraphics[scale=.90,angle=-0]{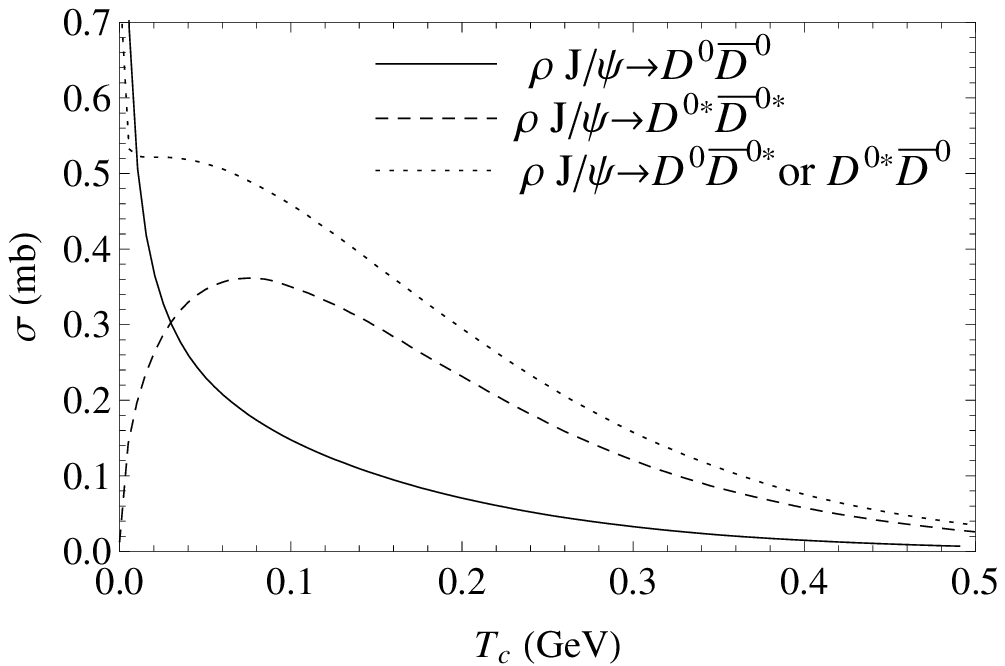}
\caption{Graph between Cross sections and $T_c$ for Quadratic
potential with Gaussian form of $f$.} \label{quadraticf}
\end{figure}

\begin{figure}[!h]
\includegraphics[scale=.90,angle=-0]{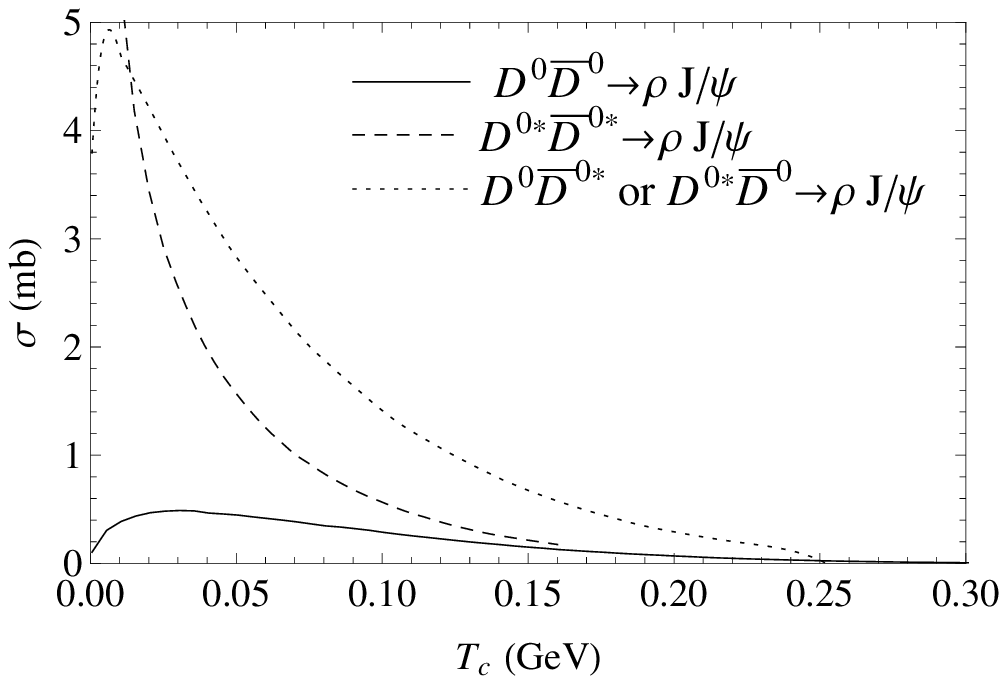}
\caption{Graph between Cross sections and $T_c$ for Cornell
potential with $f=1$ (reverse processes).} \label{reversecornell}
\end{figure}

\begin{figure}[!h]
\includegraphics[scale=.90,angle=-0]{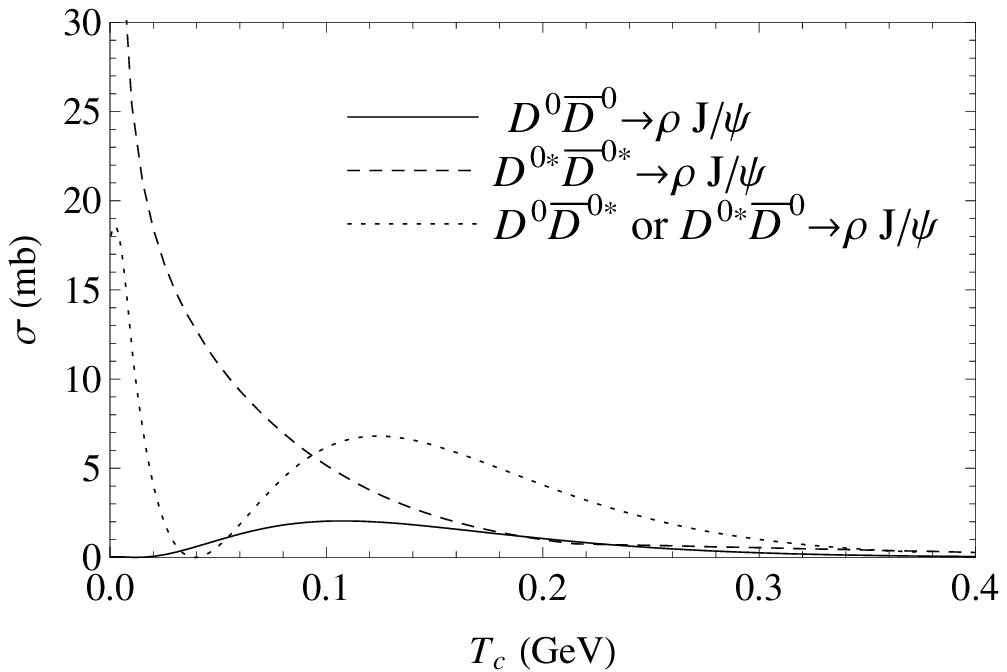}
\caption{Graph between Cross sections and $T_c$ for Quadratic
potential with $f=1$ (reverse processes).} \label{reversequadratic}
\end{figure}

\begin{figure}[!h]
\includegraphics[scale=.90,angle=-0]{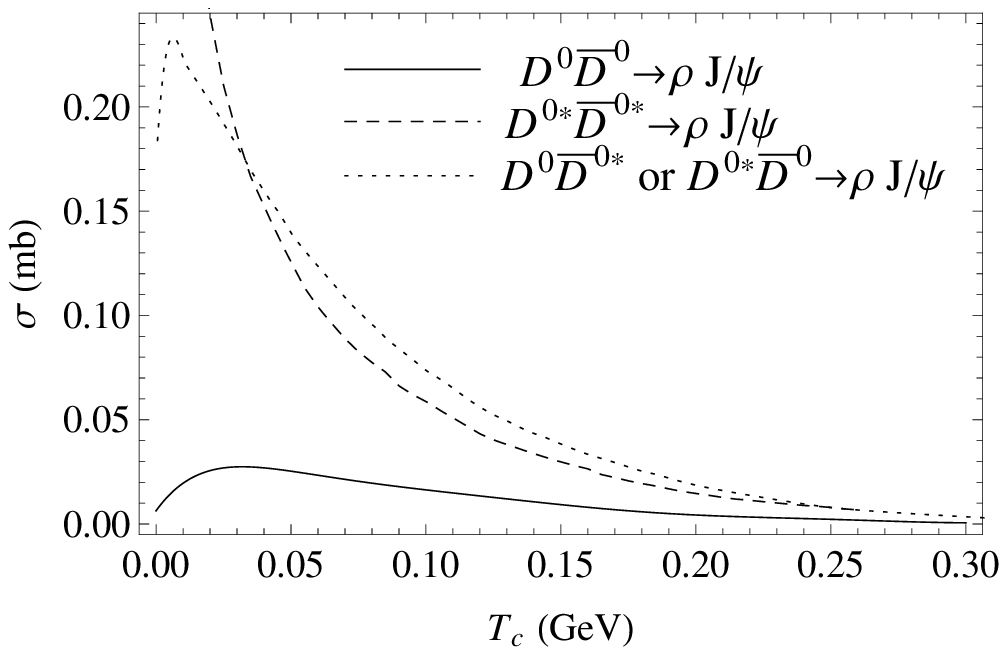}
\caption{Graph between Cross sections and $T_c$ for Cornell
potential with Gaussian form of $f$ (reverse processes).}
\label{reversecornellf}
\end{figure}

\begin{figure}[!h]
\includegraphics[scale=.90,angle=-0]{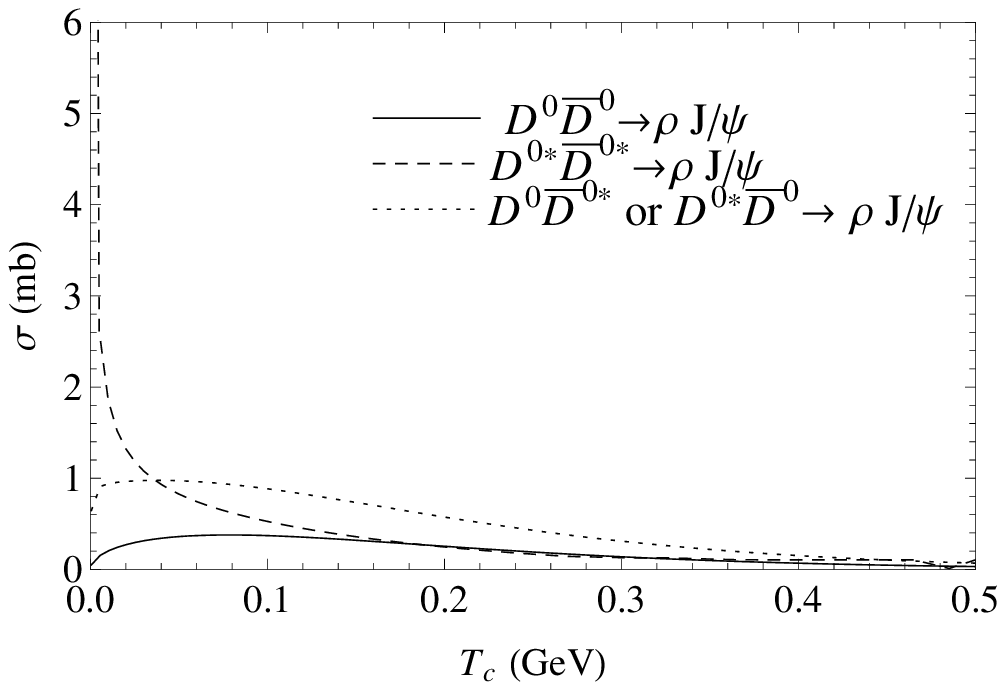}
\caption{Graph between Cross sections and $T_c$ for Quadratic
potential with Gaussian form of $f$ (reverse processes).}
\label{reversequadraticf}
\end{figure}

The values of cross sections of $\rho J/\psi\rightarrow
D^0\bar{D}^0$, $\rho J/\psi\rightarrow D^0\bar{D}^{0*}$
($D^{0*}\bar{D}^{0}$), $\rho J/\psi\rightarrow D^{0*}\bar{D}^{0*}$
and corresponding inverse processes by using Cornell as well as
quadratic potentials are given in figures (\ref{cornell}) to
(\ref{reversequadraticf}). These cross sections are first obtained
by using $f=1$ and latter by using Gaussian form of $f$. In these
figures $T_c=E_c-m_{D^{0*}}-m_{\bar D^{0*}}$ for the processes $\rho
J/\psi\rightarrow D^{0*}\bar D^{0*}$ and $D^{0*}\bar
D^{0*}\rightarrow\rho J/\psi$ and $T_c=E_c-m_{\rho}-m_{J/\psi}$ for
remaining four processes, where $E_c$ is total centre of mass
energy. The processes $\rho J/\psi\rightarrow D^0 \bar D^0$, $\rho
J/\psi\rightarrow D^0 \bar D^{0*}/D^{0*} \bar D^{0}$ and
$D^{0*}\bar{D}^{0*}\rightarrow \rho J/\psi$ are exothermic where as
all the remaining processes are endothermic. In all plots we can
observe that the cross sections obtained for the quadratic potential
have different shapes as compared to the Cornell potential. All the
reactions show larger cross sections for the quadratic potential
when compared with the Cornell potential.

There is a significant suppression in the cross sections obtained
when factor $f$ is introduced as compared to the cross sections
obtained from simple sum of two body approach. We can observe this
suppression easily in all endothermic reactions where the peak of
these cross sections is decreased by factor of 14 to 20 after
including the factor $f$. For the process $\rho J/\psi\rightarrow
D^0\bar D^{0*}$ or $D^{0*}\bar{D}^0$ there is 18 times suppression
in the cross sections for the Cornell potential after including $f$
where as for quadratic potential this suppression is 14 times. After
including $f$ the suppressions in the cross sections for the process
$D^{0}\bar{D}^0\rightarrow \rho J/\psi$ are nearly 16 and 7 times
for the Cornell and the quadratic potentials respectively and for
the process $D^0\bar D^{0*}$ or $D^{0*}\bar{D}^0\rightarrow \rho
J/\psi$ these suppressions are 20 and 18 times for the Cornell and
the quadratic potentials respectively. It is noted that the Gaussian
form of $f$ factor produces larger suppression in case of Cornell as
compared to quadratic potential. To elucidate this effect we produce
the plots of potential energy and remaining terms which include
kinetic energy and constant terms verses $T_{c}$ for a specific
process $\rho J/\psi\rightarrow D^0\bar{D}^0$. The resultant values
of cross sections depend on delicate cancelation in these terms. The
Fig. 10 shows that in case of Cornell potential these terms are
modified by the $f$ factor in such a way that cancelation is higher
as compared to quadratic potential for which the similar plot is
given in Fig. 11.

Kevin L. Haglin and Charles Gale in ref. \cite{kevin} used chiral
Lagrangian approach for heavy light mesons to study the cross
sections of different $J/\psi$ reactions. The peak of the cross
section for the process $\rho J/\psi\rightarrow D^*\bar D^*$ near
threshold is 5 mb in figure (3) of ref. \cite{kevin} which is
comparable with the dashed curve of our figure (\ref{quadratic})
where we are using quadratic potential and $f$=1. In figure (4)
Haglin et. al. also reported the cross sections for the process
$\rho J/\psi\rightarrow D\bar D$. Here cross section is maximum near
threshold and decreases to zero at total centre of mass energy $4.2$
GeV. These cross sections are comparable with the solid curve of our
figure (\ref{quadraticf}) where we are using quadratic potential and
Gaussian form of $f$. The other processes reported in our paper are
not calculated by Haglin et. al.

 Ziwei Lin and C. M. Ko in ref.
\cite{Ziwei} used an effective hadronic Lagrangian to obtain the
cross sections for different $J/\psi$ reactions. In figure (4) of
this reference, Ziwei et. al. reported the cross sections for the
process $\rho J/\psi\rightarrow D^*\bar D^*$ and $\rho
J/\psi\rightarrow D\bar D$. In this figure the peak of the cross
section for the process $\rho J/\psi\rightarrow D^*\bar D^*$ is 9 mb
near threshold without form factor which is nearly double than that
of the cross section obtained in our figure (\ref{quadratic}) when
we are using quadratic potential with $f$=1. But when they
introduced a form factor then the peak of the cross section becomes
approximately 5 mb which agrees with our result for quadratic
potential as in figure (\ref{quadratic}). Yongseok Oh, Taesoo Song,
and Su Houng Lee in ref. \cite{Yongseok} also used effective
Lagrangian for the $J/\psi$ absorption by $\pi$ and $\rho$ mesons.
In figure (6) Yongseok et. al. reported the crosss sections for the
processes $\rho J/\psi\rightarrow D\bar D$, $\rho J/\psi\rightarrow
D^*\bar D (D\bar D^*$) and $\rho J/\psi\rightarrow D^*\bar D^*$. The
shape of their plots is different with our plots except for the
process $\rho J/\psi\rightarrow D\bar D$ which damps at total centre
of mass energy $4.08$ GeV corresponding to $T_c=0.2$ GeV in
agreement with our result obtained using the Cornell potential. The
effective Lagrangian approach used in above mentioned refs.
\cite{Ziwei,Yongseok} is some what unrealistic as it is based on
using SU(4) flavour symmetry which is badly broken in QCD due to
large charm quark mass as compared to light quarks.

In ref. \cite{Wong Swanson Barnes 2000} Cheuk-Yin Wong, E. S.
Swanson and T. Barnes evaluated the cross sections for different
processes using the quark interchange model of Barnes and Swanson
\cite{Barnes1992}. In figure (3) of this ref. \cite{Wong Swanson
Barnes 2000} Wong et. al. reported the cross sections for the
processes $\rho J/\psi\rightarrow D\bar D$, $\rho J/\psi\rightarrow
D^*\bar D$ or $D\bar D^*$ and $\rho J/\psi\rightarrow D^*\bar D^*$.
Their cross section for the process $\rho J/\psi\rightarrow D\bar D$
damped at $T_c$ nearly equal to 0.5 GeV whereas in our figures
(\ref{cornell}) and (\ref{cornellf}) it is damped at $T_c=0.2$ GeV
for Cornell Potential with $f$=1 and in our figure
(\ref{quadraticf}) the cross section is damped for the same process
at $T_c$ nearly equal to 0.5 GeV when we use quadratic potential
along with Gaussian form of $f$. However this damping is at
$T_c=0.2$ GeV in their later work \cite{Wong Swanson Barnes 2001}.
In these papers Wong et. al. reported the cross sections for the
process $\rho J/\psi\rightarrow D^*\bar D^*$ with total spin 0, 1
and 2 separately whereas we are reporting spin averaged cross
sections. Wong et. al. in their approach used the Cornell potential
but not included $f$. We are reporting our cross sections after
improving the potential by including $f$. In figures (5) and (6) of
ref. \cite{Imran Jameel 2017} the cross sections are reported for
the process $\rho J/\psi\rightarrow D^0\bar{D}^{0*}$ and
$D^0\bar{D}^{0*}\rightarrow \rho J/\psi$. In this reference only
quadratic potential is used with $f=1$ and Gaussian form of $f$. In
our current work we are using broader spectrum to fit the parameters
which cause a little change in these cross sections with this given
reference. Moreover, in our recent work we are replacing quadratic
potential by more realistic Cornell potential while keep using $f$=1
and Gaussian form of $f$ as well.


\begin{figure}[!h]
\includegraphics[scale=.75,angle=-0]{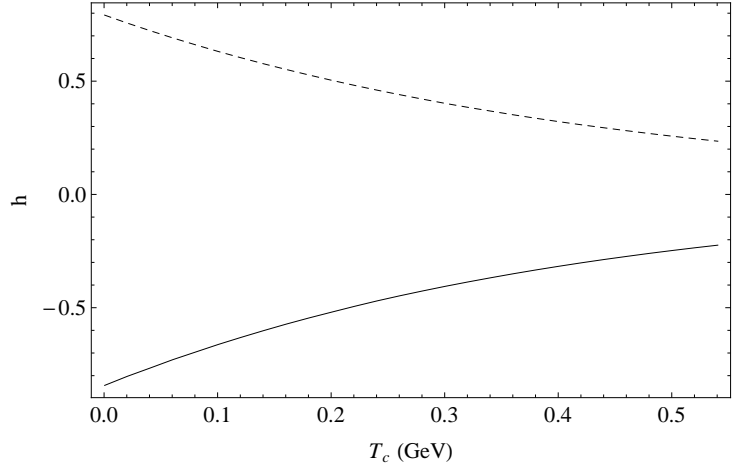}
\caption{Comparison of PE part and remaining terms of integratal
equations with Cornell Potential after including $f$ solid curve
shows potential part. } \label{Cornell comparison of KE PE}
\end{figure}
\begin{figure}[!h]
\includegraphics[scale=.70,angle=-0]{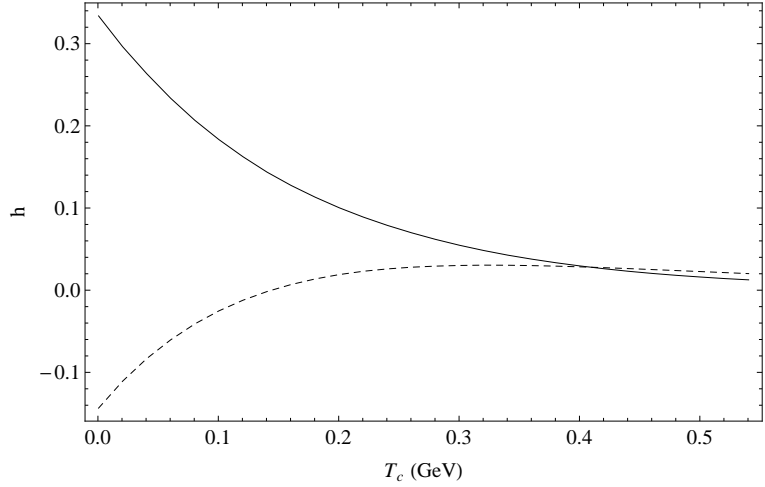}
\caption{Comparison of PE part and remaining terms of integral
equations with Quadratic Potential after including $f$. }
\label{Quadratic comparison of KE PE}
\end{figure}

~~~~~~~~~~~~~~~~~~~~~~~~~~~~~~~~~~~~~~~~~~~~~~~~~~~~~~~~~~~~~~~~~~~~~~~~~~~~~~~~~~~~~~~~~~~~~~~~
~~~~~~~~~~~~~~~~~~~~~~~~~~~~~~~~~~~~~~~~~~~~~~~~~~~~~~~~~~~~~~~~~~~~~~~~~~~~~~~~~~~~~~~~~~~~~~~~

\newpage
\begin{appendix}

\section{The spin basis}\label{spin}

For total spin 0 we have the following spin states
\begin{eqnarray}
|P_{1\bar{3}}P_{2\bar{4}}\rangle_0=\frac{1}{2}
[\uparrow\uparrow\downarrow\downarrow-\uparrow\downarrow\downarrow\uparrow-
\downarrow\uparrow\uparrow\downarrow+
\downarrow\downarrow\uparrow\uparrow]
\end{eqnarray}
\begin{eqnarray}
|V_{1\bar{4}}V_{2\bar{3}}\rangle_0=\sqrt{\frac{1}{12}}
[2\uparrow\downarrow\downarrow\uparrow+2\downarrow\uparrow\uparrow\downarrow-
\uparrow\uparrow\downarrow\downarrow-\uparrow\downarrow\uparrow\downarrow-
\downarrow\uparrow\downarrow\uparrow-\downarrow\downarrow\uparrow\uparrow]
\end{eqnarray}
\begin{eqnarray}
|V_{1\bar{3}}V_{2\bar{4}}\rangle_0=\sqrt{\frac{1}{12}}
[2\uparrow\downarrow\uparrow\downarrow+2\downarrow\uparrow\downarrow\uparrow-
\uparrow\uparrow\downarrow\downarrow-\uparrow\downarrow\downarrow\uparrow-
\downarrow\uparrow\uparrow\downarrow-\downarrow\downarrow\uparrow\uparrow]
\end{eqnarray}

the spin overlaps are
\begin{eqnarray}
_0\langle
P_{1\bar{3}}P_{2\bar{4}}|P_{1\bar{3}}P_{2\bar{4}}\rangle_0= \
_0\langle
V_{1\bar{3}}V_{2\bar{4}}|V_{1\bar{3}}V_{2\bar{4}}\rangle_0=\
_0\langle
V_{1\bar{4}}V_{2\bar{3}}|V_{1\bar{4}}V_{2\bar{3}}\rangle_0=1
\end{eqnarray}
\begin{eqnarray}
_0\langle
P_{1\bar{3}}P_{2\bar{4}}|V_{1\bar{4}}V_{2\bar{3}}\rangle_0= \
_0\langle
V_{1\bar{4}}V_{2\bar{3}}|P_{1\bar{3}}P_{2\bar{4}}\rangle_0=\frac{-\sqrt{3}}{2}
\end{eqnarray}
\begin{eqnarray}
_0\langle
V_{1\bar{3}}V_{2\bar{4}}|V_{1\bar{4}}V_{2\bar{3}}\rangle_0=\
_0\langle
V_{1\bar{4}}V_{2\bar{3}}|V_{1\bar{3}}V_{2\bar{4}}\rangle_0=\frac{-1}{2}
\end{eqnarray}
By using the formalism
$S_i.S_j=(\frac{\sigma_1}{2})_i(\frac{\sigma_1}{2})_j+(\frac{\sigma_2}{2})_i(\frac{\sigma_2}{2})_j
+(\frac{\sigma_3}{2})_i(\frac{\sigma_3}{2})_j$ where
$(\sigma_1)_i\uparrow_i=\downarrow_i$,
$(\sigma_1)_i\downarrow_i=\uparrow_i$,
$(\sigma_2)_i\uparrow_i=i\downarrow_i$,
$(\sigma_2)_i\downarrow_i=-i\uparrow_i$,
$(\sigma_3)_i\uparrow_i=\uparrow_i$,
$(\sigma_3)_i\downarrow_i=-\downarrow_i$ $(i=1,2,\bar3,\bar4)$, the
matrix elements of $S_i.S_j$ in $|1\rangle_0$ and $|2\rangle_0$
basis were calculated to be

\begin{eqnarray}
_0\langle P_{1\bar{3}}P_{2\bar{4}}|\left(
           \begin{array}{c}
             S_1.S_2 \\
             S_1.S_{\bar{3}}\\
             S_1.S_{\bar{4}} \\
             S_2.S_{\bar{3}} \\
             S_2.S_{\bar{4}} \\
             S_{\bar{3}}.S_{\bar{4}} \\
           \end{array}
         \right)
|P_{1\bar{3}}P_{2\bar{4}}\rangle_0=\left(
            \begin{array}{c}
              0 \\
              -3/4 \\
              0 \\
              0 \\
              -3/4 \\
              0 \\
            \end{array}
          \right)
\end{eqnarray}

\begin{eqnarray}
_0\langle P_{1\bar{3}}P_{2\bar{4}}|\left(
           \begin{array}{c}
             S_1.S_2 \\
             S_1.S_{\bar{3}}\\
             S_1.S_{\bar{4}} \\
             S_2.S_{\bar{3}} \\
             S_2.S_{\bar{4}} \\
             S_{\bar{3}}.S_{\bar{4}} \\
           \end{array}
         \right)
|V_{1\bar{4}}V_{2\bar{3}}\rangle_0=\ _0\langle
V_{1\bar{4}}V_{2\bar{3}}|\left(
           \begin{array}{c}
             S_1.S_2 \\
             S_1.S_{\bar{3}}\\
             S_1.S_{\bar{4}} \\
             S_2.S_{\bar{3}} \\
             S_2.S_{\bar{4}} \\
             S_{\bar{3}}.S_{\bar{4}} \\
           \end{array}
         \right)
|P_{1\bar{3}}P_{2\bar{4}}\rangle_0=\left(
            \begin{array}{c}
              \sqrt{3}/8 \\
              3\sqrt{3}/8 \\
              -\sqrt{3}/8 \\
              -\sqrt{3}/8 \\
              3\sqrt{3}/8 \\
              \sqrt{3}/8 \\
            \end{array}
          \right)
\end{eqnarray}

\begin{eqnarray}
_0\langle V_{1\bar{4}}V_{2\bar{3}}|\left(
           \begin{array}{c}
             S_1.S_2 \\
             S_1.S_{\bar{3}}\\
             S_1.S_{\bar{4}} \\
             S_2.S_{\bar{3}} \\
             S_2.S_{\bar{4}} \\
             S_{\bar{3}}.S_{\bar{4}} \\
           \end{array}
         \right)
|V_{1\bar{4}}V_{2\bar{3}}\rangle_0=\left(
            \begin{array}{c}
              -1/2 \\
              -1/2 \\
              1/4 \\
              1/4 \\
              -1/2 \\
              -1/2 \\
            \end{array}
          \right)
\end{eqnarray}

\begin{eqnarray}
_0\langle V_{1\bar{3}}V_{2\bar{4}}|\left(
           \begin{array}{c}
             S_1.S_2 \\
             S_1.S_{\bar{3}}\\
             S_1.S_{\bar{4}} \\
             S_2.S_{\bar{3}} \\
             S_2.S_{\bar{4}} \\
             S_{\bar{3}}.S_{\bar{4}} \\
           \end{array}
         \right)
|V_{1\bar{3}}V_{2\bar{4}}\rangle_0=\left(
            \begin{array}{c}
              -1/2 \\
              1/4 \\
             -1/2 \\
              -1/2 \\
              1/4 \\
              -1/2 \\ \\
            \end{array}
          \right)
\end{eqnarray}

\begin{eqnarray}
_0\langle V_{1\bar{3}}V_{2\bar{4}}|\left(
           \begin{array}{c}
             S_1.S_2 \\
             S_1.S_{\bar{3}}\\
             S_1.S_{\bar{4}} \\
             S_2.S_{\bar{3}} \\
             S_2.S_{\bar{4}} \\
             S_{\bar{3}}.S_{\bar{4}} \\
           \end{array}
         \right)
|V_{1\bar{4}}V_{2\bar{3}}\rangle_0=\ _0\langle
V_{1\bar{4}}V_{2\bar{3}}|\left(
           \begin{array}{c}
             S_1.S_2 \\
             S_1.S_{\bar{3}}\\
             S_1.S_{\bar{4}} \\
             S_2.S_{\bar{3}} \\
             S_2.S_{\bar{4}} \\
             S_{\bar{3}}.S_{\bar{4}} \\
           \end{array}
         \right)
|V_{1\bar{3}}V_{2\bar{4}}\rangle_0=\left(
            \begin{array}{c}
              5/8 \\
              -1/8 \\
              -1/8 \\
              -1/8 \\
              -1/8 \\
              5/8 \\
            \end{array}
          \right)
\end{eqnarray}
For total spin 1 we have following spin states
\begin{eqnarray}
|P_{1\bar{3}}V_{2\bar{4}}\rangle_1=\frac{1}{\sqrt{6}}[\uparrow\uparrow\downarrow\uparrow
-\downarrow\uparrow\uparrow\uparrow+\frac{1}{\sqrt{2}}(\uparrow\uparrow\downarrow\downarrow
+\uparrow\downarrow\downarrow\uparrow-\downarrow\uparrow\uparrow\downarrow-\downarrow\downarrow\uparrow\uparrow)
+\uparrow\downarrow\downarrow\downarrow-\downarrow\downarrow\uparrow\downarrow]
\end{eqnarray}
\begin{eqnarray}
|V_{1\bar{3}}V_{2\bar{4}}\rangle_1=\frac{1}{\sqrt{12}}[\uparrow\uparrow\uparrow\downarrow+\uparrow\downarrow\uparrow\uparrow
-\uparrow\uparrow\downarrow\uparrow-\downarrow\uparrow\uparrow\uparrow+\uparrow\downarrow\downarrow\downarrow
+\downarrow\downarrow\uparrow\downarrow-\downarrow\uparrow\downarrow\downarrow-\downarrow\downarrow\downarrow\uparrow\nonumber\\
+\sqrt{2}\uparrow\downarrow\uparrow\downarrow+\sqrt{2}\downarrow\uparrow\downarrow\uparrow]
\end{eqnarray}
\begin{eqnarray}
|V_{1\bar{4}}V_{2\bar{3}}\rangle_1=\frac{1}{\sqrt{12}}[\uparrow\uparrow\downarrow\uparrow
+\uparrow\downarrow\uparrow\uparrow-\uparrow\uparrow\uparrow\downarrow-\downarrow\uparrow\uparrow\uparrow
+\sqrt{2}\uparrow\downarrow\downarrow\uparrow-\sqrt{2}\downarrow\uparrow\uparrow\downarrow
+\uparrow\downarrow\downarrow\downarrow\nonumber\\
+\downarrow\downarrow\downarrow\uparrow-\downarrow\uparrow\downarrow\downarrow
-\downarrow\downarrow\uparrow\downarrow]
\end{eqnarray}
the spin overlaps are
\begin{eqnarray}
_1\langle
P_{1\bar{3}}V_{2\bar{4}}|P_{1\bar{3}}V_{2\bar{4}}\rangle_1=\
_1\langle
V_{1\bar{3}}V_{2\bar{4}}|V_{1\bar{3}}V_{2\bar{4}}\rangle_1=\
_1\langle
V_{1\bar{4}}V_{2\bar{3}}|V_{1\bar{4}}V_{2\bar{3}}\rangle_1=1
\end{eqnarray}
\begin{eqnarray}
_1\langle
P_{1\bar{3}}V_{2\bar{4}}|V_{1\bar{4}}V_{2\bar{3}}\rangle_1=\
_1\langle V_{1\bar{4}}V_{2\bar{3}}|
P_{1\bar{3}}V_{2\bar{4}}\rangle_1=\frac{1}{\sqrt{2}}
\end{eqnarray}
\begin{eqnarray}
_1\langle
V_{1\bar{3}}V_{2\bar{4}}|V_{1\bar{4}}V_{2\bar{3}}\rangle_1=0
\end{eqnarray}
The matrix elements of $S_i.S_j$ in $|1\rangle_1$ and $|2\rangle_1$
basis were calculated to be
\begin{eqnarray}
_1\langle P_{1\bar{3}}V_{2\bar{4}}|\left(
           \begin{array}{c}
             S_1.S_2 \\
             S_1.S_{\bar{3}}\\
             S_1.S_{\bar{4}} \\
             S_2.S_{\bar{3}} \\
             S_2.S_{\bar{4}} \\
             S_{\bar{3}}.S_{\bar{4}} \\
           \end{array}
         \right)
|P_{1\bar{3}}V_{2\bar{4}}\rangle_1=\left(
            \begin{array}{c}
              0 \\
              -3/4 \\
              0 \\
              0 \\
              1/4 \\
              0 \\
            \end{array}
          \right)
\end{eqnarray}
\begin{eqnarray}
_1\langle P_{1\bar{3}}V_{2\bar{4}}|\left(
           \begin{array}{c}
             S_1.S_2 \\
             S_1.S_{\bar{3}}\\
             S_1.S_{\bar{4}} \\
             S_2.S_{\bar{3}} \\
             S_2.S_{\bar{4}} \\
             S_{\bar{3}}.S_{\bar{4}} \\
           \end{array}
         \right)
|V_{1\bar{4}}V_{2\bar{3}}\rangle_1=\ _1\langle
V_{1\bar{4}}V_{2\bar{3}}|\left(
           \begin{array}{c}
             S_1.S_2 \\
             S_1.S_{\bar{3}}\\
             S_1.S_{\bar{4}} \\
             S_2.S_{\bar{3}} \\
             S_2.S_{\bar{4}} \\
             S_{\bar{3}}.S_{\bar{4}} \\
           \end{array}
         \right)
|P_{1\bar{3}}V_{2\bar{4}}\rangle_1=\left(
            \begin{array}{c}
              -1/4\sqrt{2} \\
              -3/4\sqrt{2} \\
              1/4\sqrt{2} \\
              1/4\sqrt{2} \\
              1/4\sqrt{2} \\
              -1/4\sqrt{2} \\
            \end{array}
          \right)
\end{eqnarray}
\begin{eqnarray}
_1\langle V_{1\bar{4}}V_{2\bar{3}}|\left(
           \begin{array}{c}
             S_1.S_2 \\
             S_1.S_{\bar{3}}\\
             S_1.S_{\bar{4}} \\
             S_2.S_{\bar{3}} \\
             S_2.S_{\bar{4}} \\
             S_{\bar{3}}.S_{\bar{4}} \\
           \end{array}
         \right)
|V_{1\bar{4}}V_{2\bar{3}}\rangle_1=\left(
            \begin{array}{c}
              -1/4 \\
              -1/4 \\
              1/4 \\
              1/4 \\
              -1/4 \\
              -1/4 \\
            \end{array}
          \right)
\end{eqnarray}
For total spin 2 we have following spin states
\begin{eqnarray}
|V_{1\bar{3}}V_{2\bar{4}}\rangle_2=|V_{1\bar{4}}V_{2\bar{3}}\rangle_2=\sqrt{\frac{1}{5}}[\downarrow\downarrow\downarrow\downarrow
+\uparrow\uparrow\uparrow\uparrow+\frac{1}{2}(\downarrow\uparrow\downarrow\downarrow+\downarrow\downarrow\downarrow\uparrow
+\uparrow\downarrow\downarrow\downarrow+\downarrow\downarrow\uparrow\downarrow+\uparrow\uparrow\uparrow\downarrow
+\uparrow\downarrow\uparrow\uparrow\nonumber\\
+\uparrow\uparrow\downarrow\uparrow+\downarrow\uparrow\uparrow\uparrow)
+\sqrt{\frac{1}{6}}(\uparrow\uparrow\downarrow\downarrow+\uparrow\downarrow\downarrow\uparrow+\downarrow\uparrow\uparrow\downarrow
+\downarrow\downarrow\uparrow\uparrow+\downarrow\uparrow\downarrow\uparrow+\uparrow\downarrow\uparrow\downarrow)]\nonumber\\
\end{eqnarray}
the spin overlaps are
\begin{eqnarray}
_2\langle
V_{1\bar{3}}V_{2\bar{4}}|V_{1\bar{4}}V_{2\bar{3}}\rangle_2=\
_2\langle
V_{1\bar{3}}V_{2\bar{4}}|V_{1\bar{3}}V_{2\bar{4}}\rangle_2=\
_2\langle
V_{1\bar{4}}V_{2\bar{3}}|V_{1\bar{4}}V_{2\bar{3}}\rangle_2=1
\end{eqnarray}
The matrix elements of $S_i.S_j$ in $|1\rangle_2$ and $|2\rangle_2$
basis were calculated to be
\begin{equation}
_2\langle V_{1\bar{3}}V_{2\bar{4}}|\left(
           \begin{array}{c}
             S_1.S_2 \\
             S_1.S_{\bar{3}}\\
             S_1.S_{\bar{4}} \\
             S_2.S_{\bar{3}} \\
             S_2.S_{\bar{4}} \\
             S_{\bar{3}}.S_{\bar{4}} \\
           \end{array}
         \right)
|V_{1\bar{3}}V_{2\bar{4}}\rangle_2=\ _2\langle
V_{1\bar{4}}V_{2\bar{3}}|\left(
           \begin{array}{c}
             S_1.S_2 \\
             S_1.S_{\bar{3}}\\
             S_1.S_{\bar{4}} \\
             S_2.S_{\bar{3}} \\
             S_2.S_{\bar{4}} \\
             S_{\bar{3}}.S_{\bar{4}} \\
           \end{array}
         \right)
|V_{1\bar{4}}V_{2\bar{3}}\rangle_2=\left(
            \begin{array}{c}
              1/4 \\
              1/4 \\
              1/4 \\
              1/4 \\
              1/4 \\
              1/4 \\
            \end{array}
          \right)
\end{equation}
\begin{equation}
_2\langle V_{1\bar{3}}V_{2\bar{4}}|\left(
           \begin{array}{c}
             S_1.S_2 \\
             S_1.S_{\bar{3}}\\
             S_1.S_{\bar{4}} \\
             S_2.S_{\bar{3}} \\
             S_2.S_{\bar{4}} \\
             S_{\bar{3}}.S_{\bar{4}} \\
           \end{array}
         \right)
|V_{1\bar{4}}V_{2\bar{3}}\rangle_2=\ _2\langle
V_{1\bar{4}}V_{2\bar{3}}|\left(
           \begin{array}{c}
             S_1.S_2 \\
             S_1.S_{\bar{3}}\\
             S_1.S_{\bar{4}} \\
             S_2.S_{\bar{3}} \\
             S_2.S_{\bar{4}} \\
             S_{\bar{3}}.S_{\bar{4}} \\
           \end{array}
         \right)
|V_{1\bar{3}}V_{2\bar{4}}\rangle_2=\left(
            \begin{array}{c}
              1/4 \\
              1/4 \\
              1/4 \\
              1/4 \\
              1/4 \\
              1/4 \\
            \end{array}
          \right)
\end{equation}

\end{appendix}

\end{document}